\documentclass{INCOTHERM2025}

\ifpdf \usepackage[pdftex]{graphicx} \pdfcompresslevel=9
\else 
\usepackage[dvips]{graphicx} \fi

\PrintedOrElectronic

\usepackage{t1enc,dfadobe}

\usepackage{egweblnk}

\usepackage{apacite}
\usepackage{amsmath}
\usepackage{stfloats}
\usepackage{siunitx}

\title
{Direct Numerical Simulation of Transonic Flows Induced Pitching of NACA Airfoil}

\author[Chandan Kumar Bhardwaj \& Swagata Bhaumik]
{\parbox{\textwidth}{\centering Chandan Kumar Bhardwaj$^{1*}$
		and Swagata Bhaumik$^{2}$ 
	}
	\\
	{\parbox{\textwidth}{\centering $^1$Department of Mechanical Engineering, IIT(ISM) Dhanbad, Dhanbad-826004, Jharkhand, India\\
			$^2$Department of Mechanical Engineering, IIT(ISM) Dhanbad, Dhanbad-826004, Jharkhand, India\\
			*22dr0073@iitism.ac.in
		}
	}
}

\begin{document}

		\maketitle	
\begin{abstract}
		In the transonic flow regime, complex phenomena like shock boundary layer interaction, shock-induced flow separation, shock buffeting, and airfoil fluttering occur. A Direct Numerical Simulation is performed to investigate the flow physics over a \textup{NACA 0012} airfoil under transonic flow conditions. The study combines fluid dynamics with structural dynamics, attempting to solve a single degree of freedom \textup{(DOF)} equation of motion to determine the pitching frequency. Analysis is performed for angle of attack $(\alpha)$ $8^{\circ}$ with varying reduced velocity $(U^*)$ from $2$ to $8$, reduced mass $(m^*)$ $5.0$, Reynolds number $(Re_{\infty})$ $3\times10^6$, and Mach number $(M_{\infty})$ $0.8$. The speed of sound in air is calculated as $a = \sqrt{\gamma R T_{\infty}}$. It is noted that the magnitude of the aerodynamic loads and moments increased with higher reduced velocity. The Fast Fourier Transform \textup{(FFT)} of the moment coefficient ($C_M$) in the fixed-airfoil condition is carried out to extract the non-dimensional dominant frequency, referred to as the Strouhal number ($St$), along with the coupled oscillation frequency of the airfoil ($St_c$). The FFT analysis for the stationary case at $\alpha = 8^{\circ}$ yields a value of $St = 1.3161$. The presence of high-frequency components indicates the onset of instability associated with shock-boundary layer interaction. The computed values of the coupled oscillation frequency ($St_c$) are $0.0762$ and $0.1716$ for the cases with angle of attack \textup{(AoA)} $8^{\circ}$ and reduced velocities $U^*=2.0$ and $8.0$, respectively. The corresponding natural aeroelastic frequencies ($St_n$) are taken as $1/U^*$, yielding values of $0.5$ and $0.125$. It is observed that for $U^* = 2$, the coupled frequency ($St_c$) aligns with the subharmonic of shock-buffet frequency, indicating flow-dominated dynamics. In contrast, for $U^* = 8$, $St_c$ coincides with the natural pitching frequency, indicating a dominant influence of the structural mode. Pressure fluctuation contours are examined to analyze the characteristics of acoustic radiation. Vorticity and schlieren visualizations are employed to identify the nature of vortex structures, the strength of the shock, and its spatial location, respectively. 

\end{abstract}
		
\section{Introduction}
	The transonic flow regime is critically important from an aerodynamic and flight performance perspective. A wide range of aircraft, including commercial jets, UAVs, and defense aircraft, operate within this regime. It is characterized by complex phenomena such as shock-boundary layer interaction (SBLI), shock-induced flow separation, and shock-induced buffeting. The occurrence of these phenomena gives rise to unsteady aerodynamic forces and moments acting on the airfoil, leading to high-frequency oscillations commonly referred to as flutter. Flutter can significantly reduce the airfoil’s fatigue life and may damage its structure. To minimize this effect, the airfoil is allowed to pitch in response to unsteady aerodynamic forces. The interaction between the fluid dynamic mode and the structural pitching mode of the airfoil leads to a transition from flutter to a limit cycle oscillation (LCO). Studying the physics of transonic flow and its aeroelastic responses is crucial for enhancing flight control and advancing modern aircraft design. Han et al.~\cite{han2022numerical} numerically investigated transonic buffet over the stationary OAT15A airfoil and reported a high-frequency component at 2274 Hz in addition to the primary buffet frequency. The study linked the high-frequency response to coupling between K\'arm\'an vortex shedding and the shock-buffeting. Through Proper Orthogonal Decomposition (POD), they further showed that the first $10$ modes were concentrated at the shock-buffet frequency and its harmonics in pitching case. In this study, an angle of attack of ($\alpha$) $8^\circ$ is used, as at $M = 0.8$ it lies within the transonic buffet range for the NACA 0012 airfoil. The reduced mass ratio $m^* = 5.0$ is based on values commonly used in earlier aeroelastic studies \cite{bose2018dynamical}, representing a structure that is neither too light nor too heavy, allowing noticeable interaction between the airfoil and the flow. Cases with reduced velocities $U^* = 2.0$ and $U^* = 8.0$ were chosen to study how changing from a low to a high reduced velocity affects the dominant frequency of motion.
	
\section{Methodology}
	A time-accurate, high-accuracy in-house compressible flow solver is developed for solving the compressible Navier-Stokes equations. The solver employs a second-order accurate, optimized upwind scheme (OUCS3) \cite{sengupta2013high} for spatial derivatives and a fourth-order Runge-Kutta method for time integration. The viscous term is discretized using the second-order central difference explicit scheme. A pressure-based JST switch \cite{jameson1981numerical} is used to capture shocks and diffuse spurious numerical oscillations accurately.

	\subsection{Geometry and Computational Domain}
	\begin{figure}[h]
\centering 
\includegraphics[width=0.8\linewidth]{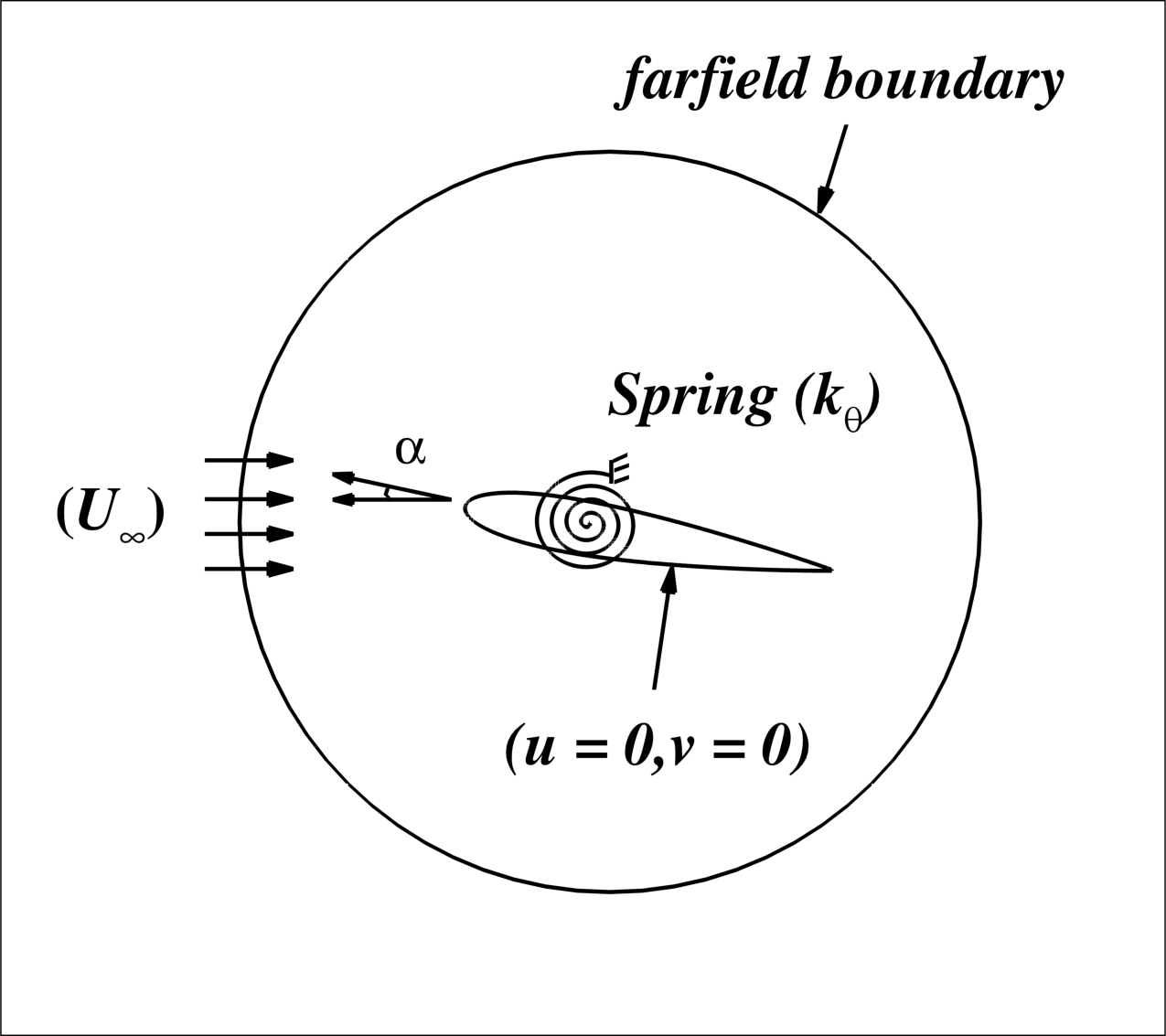}
\caption{\label{fig:1}
\textbf{Computational Domain}}
\end{figure} 

	A symmetric NACA 0012 airfoil, elastically mounted using a linear spring of stiffness $k_{\theta}$, is subjected to transonic flow conditions for numerical simulation, as depicted in Fig. \ref{fig:1}. The airfoil has a chord length of 172.85 mm. The maximum thickness of the airfoil is located at 12 percent of the chord length measured from the leading edge. The far-field boundary of the computational domain is located at a distance 60 times of the chord length (\textit{60c}) from the airfoil. This domain size is selected to minimize the influence of far-field boundary reflections on the computed flow field. The airfoil is pivoted at a point located 0.25c from the leading edge, corresponding to its aerodynamic center. This configuration allows the airfoil to pitch freely in response to unsteady aerodynamic moments.
	
	\subsection{Grid Generation}
\begin{figure}[h!]
\centering
\includegraphics[width=0.8\linewidth]{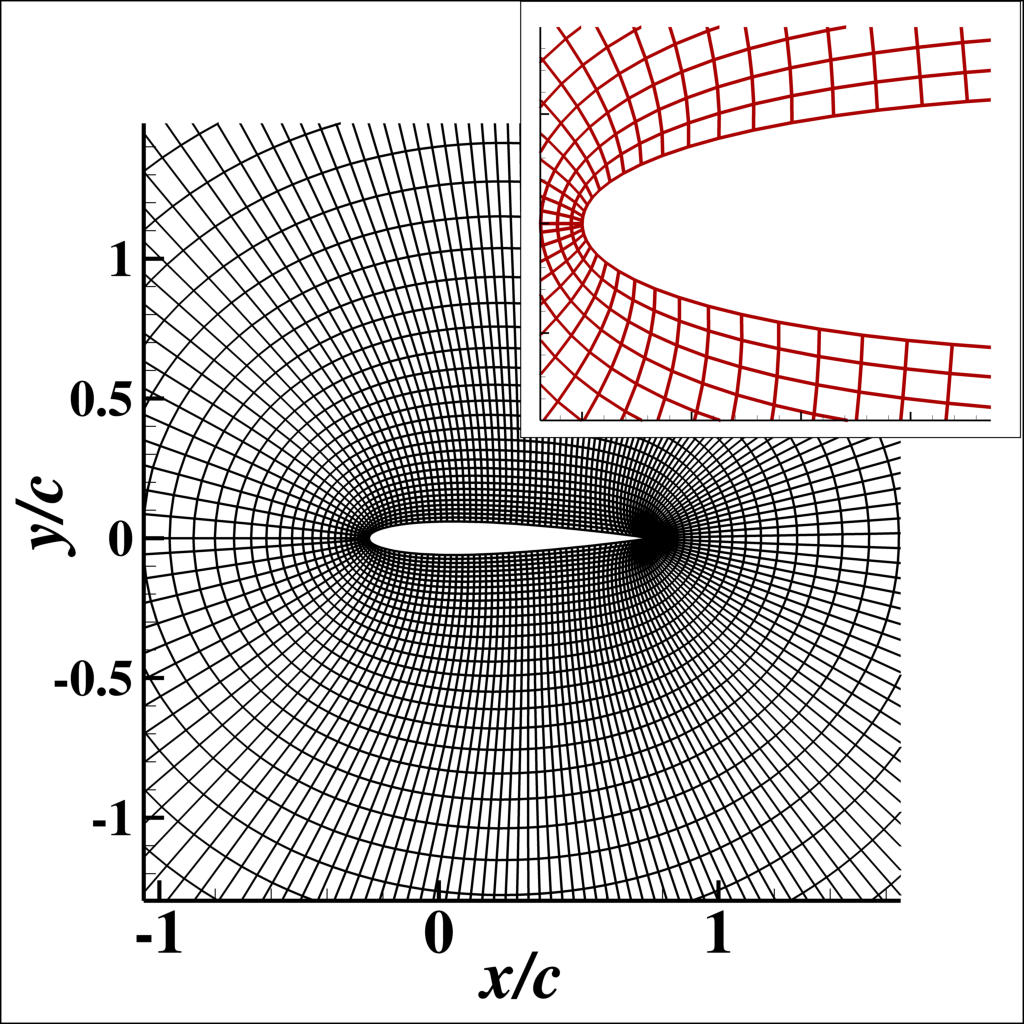}
\caption{\label{fig:2}
\textbf{Orthogonal O-Grid NACA0012}}
\end{figure}

	A structured orthogonal elliptic O-type grid is employed for the simulation. Elliptic grid generation by solving the elliptic partial differential equation using the successive over-relaxation method \cite{hoffmann2000computational} is done with 1001 points in the azimuthal direction, and 501 points in the wall-normal direction. The initial condition of the above grid generation technique was used from the solution of the hyperbolic Beltrami equation \cite{nair1998orthogonal}. The tangent hyperbolic function is used to cluster grid points close to the surface. Elliptic smoothing of the initial hyperbolic grid was done to avoid the presence of grid shock. The grid spacing at the boundary, i.e., $\Delta y_{1} = 0.002803$, and the topmost grid spacing is $\Delta y_{top} = 0.257252$. There is no grid shock in the generated grid, and the grid is almost orthogonal as shown in Fig. \ref{fig:2}. About 10${^\circ}$ deviation from orthogonality is noted very close to the trailing edge. The outer boundary of the computational domain is located at 60c, which produces an almost uniform grid in the wake region, in both directions.
	
	\subsection{Governing Equations and Boundary Conditions}
	
	The 2D compressible Navier-Stokes equations are numerically solved to simulate transonic flow. The Non-dimensional Navier-Stokes equation in strong conservative form on the transformed plane can be written as follows: 
	
\begin{equation}
\frac{\partial \hat{Q}}{\partial \tau} + \frac{\partial \hat{E}}{\partial \xi} + \frac{\partial \hat{F}}{\partial \eta} = \frac{\partial \hat{E_{v}}}{\partial \xi} + \frac{\partial \hat{F_{v}}}{\partial \eta}
\label{eq11} 
\end{equation}   
	
\(\xi\) and \(\eta\) are the curvilinear coordinate systems along azimuthal and radial directions. More detailed information about the equation can be seen in \cite{chung2002computational}. The simulation is conducted using air as the working fluid having free stream temperature, dynamic viscosity and density to be $T_{\infty} = 303$ \si{K}, $\mu_{\infty} = 1.872 \times 10^{-5}$ \si{Pa.s} and $\rho_{\infty} = 1.164$ \si{kg/m^{3}}. Prandtl number of the flow is considered to be $P_{r} = 0.7$ and \(\mu\) depends upon the fluid temperature 'T' via Sutherland's law. The Stoke's hypothesis \(\left[\lambda = -\frac{2}{3}\mu\right]\) is considered here. The time step employed in the simulation is $\Delta t = 1 \times 10^{-5}$.

The Non-dimensional single degree of freedom equation of motion that is the pitching motion equation used for solving the pitching of the airfoil is:

\begin{equation}
{\ddot\theta} + \left(\frac{2\pi}{U^*}\right)^2 \theta = \frac{C_{M}}{2m^*}\left(\frac{L_{c}}{R_g}\right)^2
\label{eq5}
\end{equation}

Where, $U^* = \frac{U_{\infty}}{f_{N}L_{c}}$, lenght scale $L_c = \frac{Re\mu}{\rho_{\infty}U_{\infty}}$ and the reduced mass $m^* = \frac{m}{\rho_{\infty}\left(A\right)}$.  A denotes the non-dimensional cross-sectional area of the airfoil. The no-slip (u = 0) and No-penetration (v = 0) conditions are applied to the airfoil surface. Riemann invariants based boundary conditions are used in the far-field boundary. To determine the surface temperature distribution, an adiabatic wall condition is imposed, implying zero heat flux across the wall. This study employs a 2D DNS, which cannot fully capture inherently 3D buffeting or flutter effects in real airfoils. Future work will extend the simulations to 3D to address this limitation.

\begin{figure*}[htb!]
\centering
\includegraphics[width=0.8\linewidth]{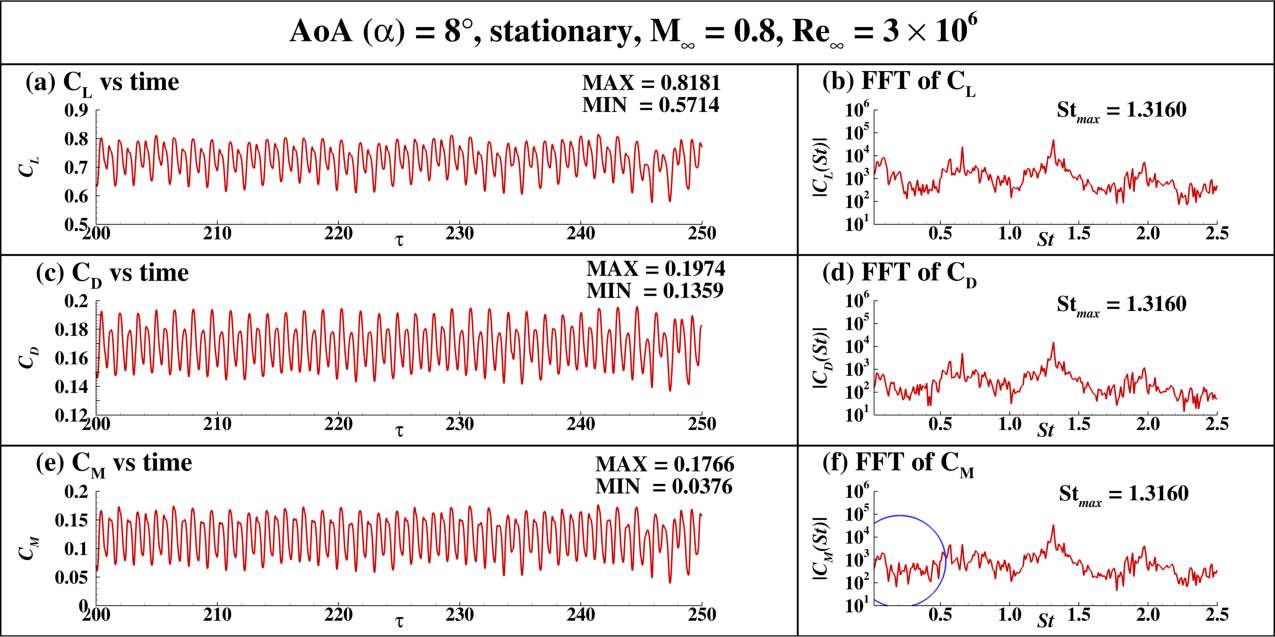}
\caption{\label{fig:AOA8_STAT_CLCDCMFFT}
\textbf{Time series and Fast Fourier Transform \textnormal{(FFT)} of lift, drag, and moment coefficients $(C_L$, $C_D$, $C_M)$ for the stationary airfoil case.}}
\end{figure*}

\section{Results and Discussion}
The nonlinear interaction between aerodynamic forces and the structural pitch mode can prevent the onset of divergent flutter in the system. Instead, the instability may saturate into a self-sustained Limit Cycle Oscillation (LCO). Fig. \ref{fig:AOA8_STAT_CLCDCMFFT} shows the time variation of $C_L$, $C_D$, and $C_M$ with their corresponding Fast Fourier Transform spectrum for the stationary case. Significant aperiodic and high-amplitude oscillations are observed in the time series data. An identical dominant Strouhal number of $St = 1.3160$ is observed in the FFT spectrum for all $C_{L}$, $C_{D}$, and $C_{M}$ time series, indicating a high-frequency chaotic flow regime. In the stationary case, this peak originates from Kelvin–Helmholtz instabilities in the separated shear layer over the airfoil. These instabilities roll up into vortices, causing the separation bubble to “breathe” at a much higher frequency than the low-frequency buffet motion. This distinct peak shows that, even without structural motion, the flow has energetic shear-layer dynamics that can influence or modulate the buffet mode. Such behavior has also been reported in previous high-fidelity simulations of transonic buffet \cite{dandois2018large}, highlighting its importance in understanding buffet mechanisms. These oscillations may result in airfoil flutter and fatigue failure.

\begin{figure*}[htb!]
\centering
\includegraphics[width=0.8\linewidth]{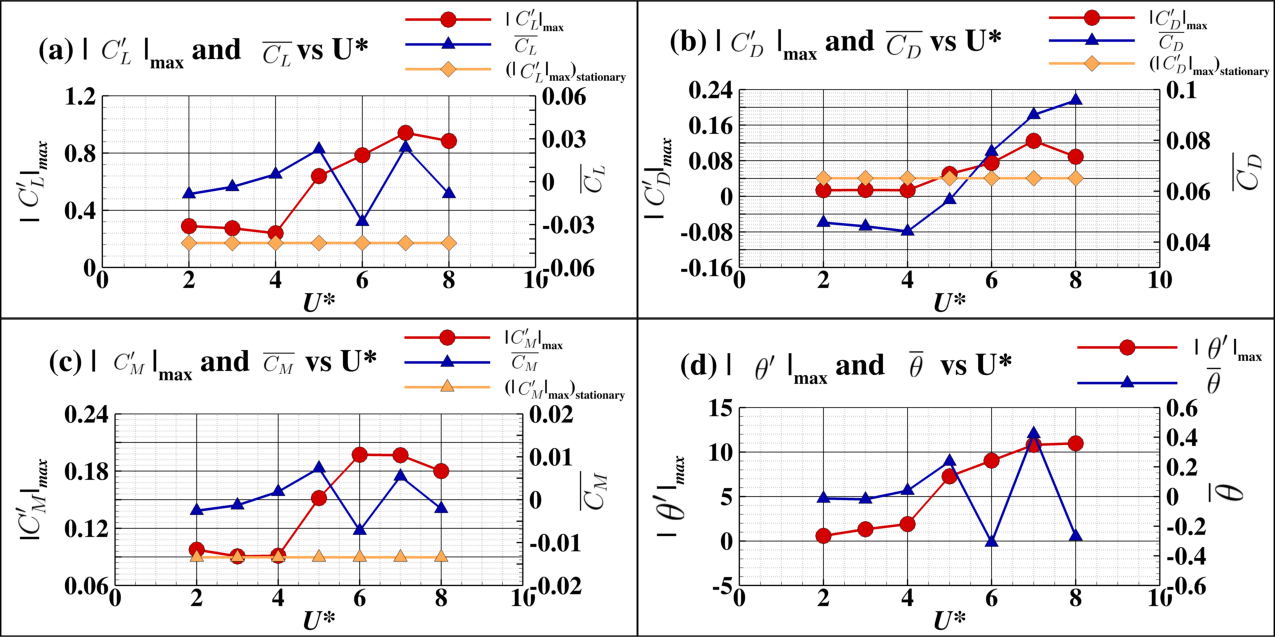}
\caption{\label{fig:3}
\textbf{Maximum magnitude of lift, drag and moment coefficient fluctuation $(|C_{L}'|_{max}$, $|C_{D}'|_{max}$, $|C_{M}'|_{max})$ along with angular displacement $(|\theta'|_{max})$ and corresponding mean values $(\overline{C_L}$, $\overline{C_D}$, $\overline{C_M}$, $\overline{\theta})$ plotted as a function of Reduced velocity $(U^*)$. Horizontal line is $|C_{L}'|_{max}$, $|C_{D}'|_{max}$ and $|C_{M}'|_{max}$ value of stationary case plotted with $U^*$.}}
\end{figure*}

\begin{figure*}[htb!]
\centering
\includegraphics[width=0.8\linewidth]{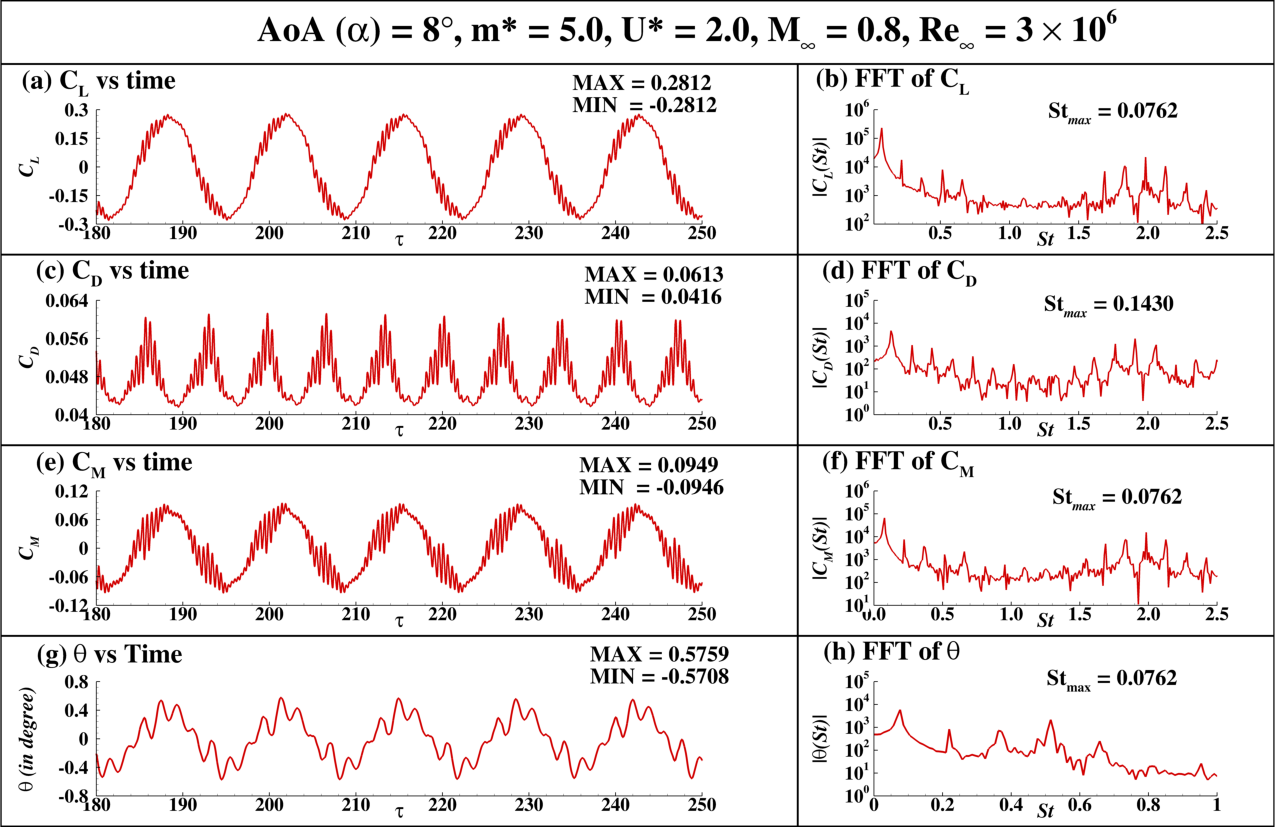}
\caption{\label{fig:AOA8_UST2_CLCDCMTHETAFFT}
\textbf{Time series and Fast Fourier Transform \textnormal{(FFT)} of lift, drag, and moment coefficients $(C_L$, $C_D$, $C_M)$ along with angular displacement $(\theta)$ for $U^* = 2.0$ case.}}
\end{figure*}

Fig. \ref{fig:3}(a) illustrates the variation of $|C_{L}'|_{max}$ with reduced velocity. It reaches its maximum value at $U^* = 7.0$. Comparison with the stationary case indicates that similar values are observed at lower $U^*$ conditions. This is expected because at a lower $U^*$, the aero-elastic pitch response is similar to the stationary airfoil case because of high spring stiffness. The airfoil behaves almost rigidly and shows weak coupling with the fluid mode. At intermediate to high $U^*$, there is a possibility of resonance coupling with the shock buffeting frequency or with its harmonics or subharmonics; at higher values of $U^*$, a chaotic regime may be encountered. This could lead to a high amplitude of $|C_{L}'|_{max}$  at higher reduced velocity. An identical trend could be noted for the drag coefficient shown in Fig. \ref{fig:3}(b). However, the value of $|C_{D}'|_{max}$  for $U^* \le 4$ is higher in the stationary case than in the pitching case, mainly because the stationary condition involves a higher effective angle of attack. Fig. \ref{fig:3}(c) and (d) depicts the variation of $|C_{M}'|_{max}$ and $|\theta'|_{max}$ with reduced velocity. Similar to the lift and drag coefficients, the maximum fluctuation magnitudes of both the moment coefficient and angular displacement exhibit comparable behavior between the elastic and stationary cases at low reduced velocity values. In the high reduced velocity regime, the maximum magnitude of $|C_{M}'|_{max}$ attains its peak at $U^* = 6.0$, beyond which it exhibits a decreasing trend. But, $|\theta'|_{max}$ reaches its maximum value at $U^* = 8.0$. This behavior results from strong coupling between the structural mode and the fluid dynamic mode when the pitching frequency approaches the natural buffeting frequency or its harmonics at $U^* = 6$. This strong aero-elastic response modifies the flow dynamics, which in turn results in a high $|C_{M}'|_{max}$ value in the $U^* =6$ case. At $U^* = 8$, the coupling between the fluid and structural modes is comparatively weaker. The couple frequency \textit{($St_{c}$)} follows the natural frequency of the elastic airfoil, resulting in  $|\theta'|_{max}$ at $U^* = 8$, despite $|C_{M}'|_{max}$ attaining its peak at $U^* = 6$.

The mean value of the coefficient of lift, drag, moment, and angular displacement is plotted as a function of $U^*$, shown in Fig. \ref{fig:3}(a to d). The $\overline{C_{L}}$, $\overline{C_{M}}$ and $\overline{\theta}$ show identical increasing trend for low $U^*$ values. As the reduced velocity increases, the spring stiffness decreases, which in turn enhances the magnitudes of the aerodynamic coefficients. At low $U^*$, weak linear coupling causes the pitch frequency to align with the harmonics or subharmonics of buffet frequency , resulting in quasi-periodic oscillations. The coupling between the two modes becomes highly nonlinear for $U^* > 4$. This occurs because the coupled frequency (\textit{$St_{c}$}) locks onto the natural elastic frequency \textit{$St_{n}$}. The amplitude of oscillation is high enough to alter the flow field, shock location, and vortex shedding due to shock-boundary interaction. Thus, for $U^* > 4$, a fluctuating trend of the mean aerodynamic coefficient is noted. The $\overline{C_{D}} $ shows an increasing trend after $U^* > 4$ as the mean drag fluctuation is dominantly dependent on shock-boundary layer interaction. Broadly, we can differentiate two zones, one for the lower ($U^* \leq 4$) case, in which the aerofoil behaves as a rigid airfoil. In a high reduced velocity zone ($U^* > 4$), in which the aerofoil may get coupled to the elastic mode. To study and analyze the physics and flow behaviour of these two zones, cases for each zone with $U^*$ = 2 and 8 are selected.
 
\begin{figure*}[htb!]
\centering
\includegraphics[width=0.8\linewidth]{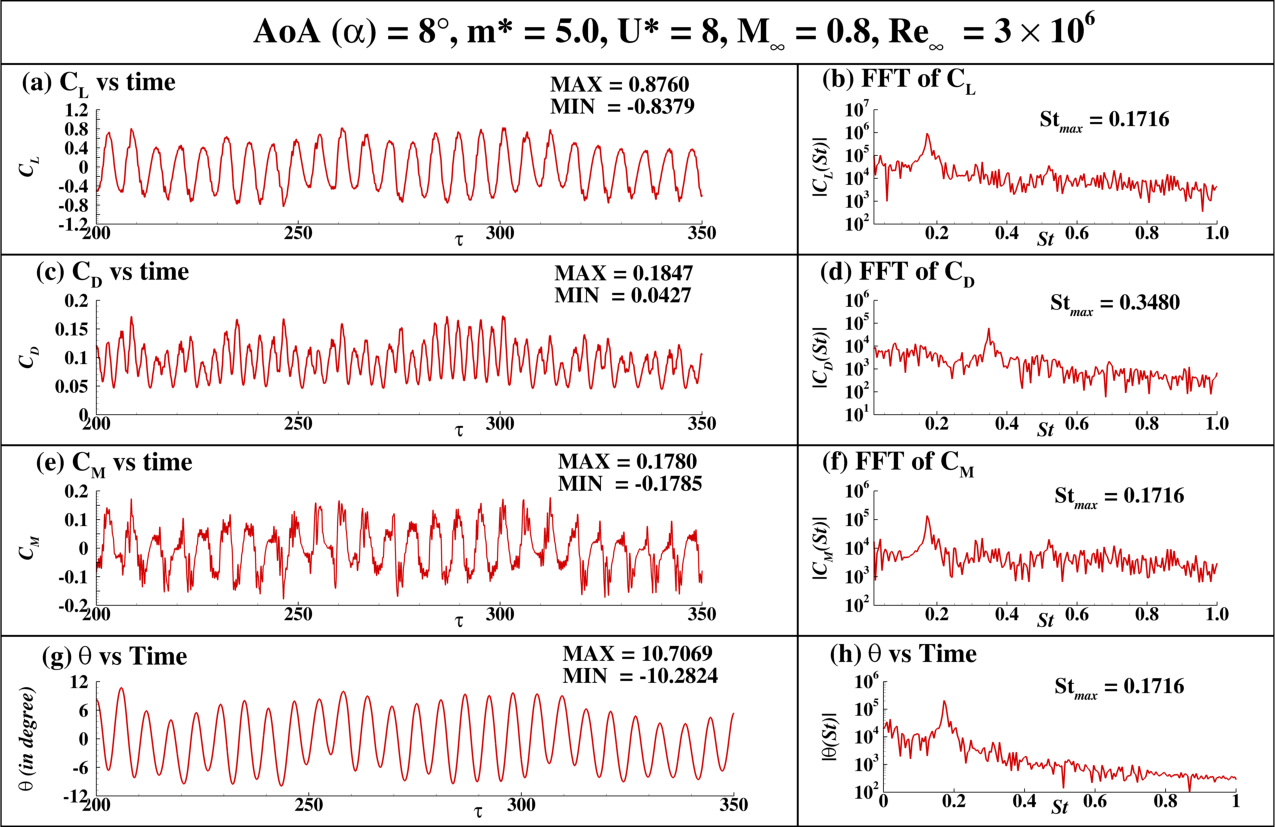}
\caption{\label{fig:AOA8_UST8_CLCDCMTHETAFFT}
\textbf{Time series and Fast Fourier Transform \textnormal{(FFT)} of lift, drag, and moment coefficients $(C_L$, $C_D$, $C_M)$ along with angular displacement $(\theta)$ for $U^* = 8.0$ case.}}
\end{figure*}

Fig. \ref{fig:AOA8_UST2_CLCDCMTHETAFFT} depicts time series of $C_{L}$, $C_{D}$, $C_{M}$ and $\theta$ with their corresponding FFt spectrum for $U^* = 2$ case. The same parameters are plotted for the $U^* = 8$ case in Fig. \ref{fig:AOA8_UST8_CLCDCMTHETAFFT}. Comparing the time series plots of lift and moment for both cases, near-periodic and highly aperiodic behavior of the series can be noted for $U^* = 2$ and 8, respectively. Fig. \ref{fig:AOA8_UST2_CLCDCMTHETAFFT}(b,f) illustrates the dominant frequency for lift and moment time series for the $U^* = 2$ case. For $U^* = 2$, multiple spectral peaks are observed that do not correspond to harmonics of the dominant frequency, indicating a quasi-periodic response. In contrast, the spectral content at $U^* = 8$ is characterized by a single dominant frequency with several broadened components, suggesting the onset of chaotic dynamics due to strong nonlinear fluid-structure interaction depicted in frame (b,f) of Fig. \ref{fig:AOA8_UST8_CLCDCMTHETAFFT}. In the periodic and quasi-periodic regimes, the oscillations generate predictable and regular aerodynamic loading on the airfoil. In contrast, the chaotic regime is characterized by a broadened frequency spectrum in the FFT, indicating frequent and irregular variations in aerodynamic loading. Such unpredictable loading can accelerate fatigue failure of the airfoil. The drag coefficient shows a distinct FFT response due to its strong dependence on boundary layer dynamics. Fig. \ref{fig:AOA8_UST2_CLCDCMTHETAFFT}(h) presents the FFT spectrum of pitching angle time series for the $U^* = 2$ case, showing a non-dimensional coupled frequency of $St_{c} = 0.07629$, while for $U^* =8$, $St_{c} = 0.1716$ shown in Fig. \ref{fig:AOA8_UST8_CLCDCMTHETAFFT}(h). This indicates that at low $U^*$, the airfoil motion aligns with the subharmonics of shock buffet frequency, whereas at high $U^*$, it follows the natural aeroelastic frequency. This subharmonic frequency is identified from the low-frequency peak in the stationary-airfoil FFT spectrum shown in Fig. \ref{fig:AOA8_STAT_CLCDCMFFT}(f) (blue circle). Its value is found to be 0.0841, which closely matches the frequency observed in the $U^* = 2$ case. For both cases, $U^* = 2.0$, and $U^* = 8.0$ the high-frequency component was not observed, unlike in the stationary case. This indicates that the coupling of the shock-buffet mode with the aeroelastic pitch mode diminishes the dominance of shear-layer instabilities.

\begin{figure*}[htb!]
\centering
\includegraphics[width=0.8\linewidth]{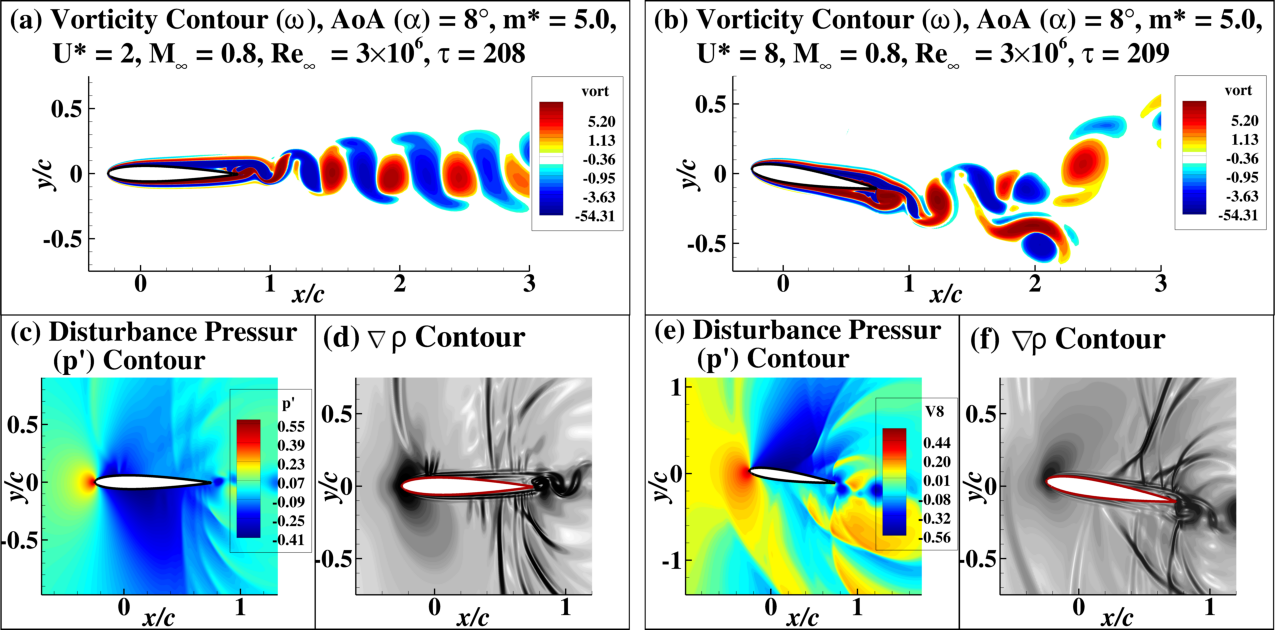}
\caption{\label{fig:AOA8_UST2_UST8_VORT_PRES_SCH}
\textbf{Vorticity $(\omega)$, Disturbance pressure \textup{(p')} and magnitude of density gradient $(|\nabla \rho|)$ contours plotted at $\tau = 208$ for $U^* = 2.0$ case and at $\tau = 209$ for $U^* = 8.0$ case.}}
\end{figure*} 

In Fig. \ref{fig:AOA8_UST2_UST8_VORT_PRES_SCH}, vorticity, pressure fluctuation, and schlieren contour of $U^* =$ 2 and 8 are shown at $\tau = 208$ and $\tau = 209$ for the respective cases. At these time instants, the airfoil attains its peak angular displacement. A 2S vortex shedding pattern is observed for the $U^* = 2$ case, whereas a P+S pattern is identified for $U^* = 8$. This distinct pattern of vorticity results from the interaction between shock-boundary layers influenced by fluid-structure coupling. Shock-associated noise radiation is much dominant for the $U^* = 8$ case depicted in Fig. \ref{fig:AOA8_UST2_UST8_VORT_PRES_SCH}(e). The Schlieren contour depicts the shock location and is present at almost the trailing edge with a boundary layer separation zone for the $U^* = 2$ case. For the $U^* = 8$ case, a $\lambda$ shock is noted to be present at the mid chord length.
	\section{Conclusions}
	
	The investigation of the complex transonic flow dynamics presented in this manuscript has led to several key insights. The main conclusions are as follows:

\begin{description}
  \item[$\bullet$] For the stationary case, the dominant frequency obtained from the FFT of the moment coefficient time series is $St = 1.316$. This high-frequency content is likely associated with vortex shedding originating from shear layer instabilities developing in the separated flow region downstream of the shock, induced by shock-boundary layer interaction.
  \item[$\bullet$] The value of $|C_{M}'|_{max}$ reaches its peak at $U^* = 6$, while the amplitude of angular fluctuation is highest at $U^* = 8$. This is possibly due to the coupling of the aeroelastic pitch frequency to a harmonic or subharmonic of the buffeting frequency at $U^* = 6$.
  \item[$\bullet$] Upon comparing the $U^* = 2$ case with the $U^* = 8$ case, it is noted that at $U^* = 2$, the aeroelastic behaviour is quasi-periodic, whereas for the $U^* = 8$ case, it exhibits chaotic behaviour.
  \item[$\bullet$] 2S and 2P+S vortex patterns are shown by reduced velocity $(U^*)$ 2 and 8 cases, respectively. The acoustic radiation domination is noted in the $U^* = 8$ case.
  
\end{description}

	
		\noindent
	\textbf{Acknowledgement}\\
	\noindent The authors gratefully acknowledge the use of HPC Aryabhatta at IIT (ISM) Dhanbad for the computational resources. The simulations and data generation in this study were made possible through its support. \\

	\bibliographystyle{apacite}
	\bibliography{bibsample}

\begin{thebibliography}{}

\bibitem [\protect \citeauthoryear {%
Bose%
, Gupta%
\BCBL {}\ \BBA {} Sarkar%
}{%
Bose%
\ \protect \BOthers {.}}{%
{\protect \APACyear {2018}}%
}]{%
bose2018dynamical}
\APACinsertmetastar {%
bose2018dynamical}%
\begin{APACrefauthors}%
Bose, C.%
, Gupta, S.%
\BCBL {}\ \BBA {} Sarkar, S.%
\end{APACrefauthors}%
\unskip\
\newblock
\APACrefYearMonthDay{2018}{}{}.
\newblock
{\BBOQ}\APACrefatitle {Dynamical behavior of unsteady flowfield of an
  elastically mounted flapping airfoil} {Dynamical behavior of unsteady
  flowfield of an elastically mounted flapping airfoil}.{\BBCQ}
\newblock
\APACjournalVolNumPages{AIAA Journal}{56}{5}{2062--2069}.
\PrintBackRefs{\CurrentBib}

\bibitem [\protect \citeauthoryear {%
Chung%
}{%
Chung%
}{%
{\protect \APACyear {2002}}%
}]{%
chung2002computational}
\APACinsertmetastar {%
chung2002computational}%
\begin{APACrefauthors}%
Chung, T.%
\end{APACrefauthors}%
\unskip\
\newblock
\APACrefYear{2002}.
\newblock
\APACrefbtitle {Computational fluid dynamics} {Computational fluid dynamics}.
\newblock
\APACaddressPublisher{}{Cambridge university press}.
\PrintBackRefs{\CurrentBib}

\bibitem [\protect \citeauthoryear {%
Dandois%
, Mary%
\BCBL {}\ \BBA {} Brion%
}{%
Dandois%
\ \protect \BOthers {.}}{%
{\protect \APACyear {2018}}%
}]{%
dandois2018large}
\APACinsertmetastar {%
dandois2018large}%
\begin{APACrefauthors}%
Dandois, J.%
, Mary, I.%
\BCBL {}\ \BBA {} Brion, V.%
\end{APACrefauthors}%
\unskip\
\newblock
\APACrefYearMonthDay{2018}{}{}.
\newblock
{\BBOQ}\APACrefatitle {Large-eddy simulation of laminar transonic buffet}
  {Large-eddy simulation of laminar transonic buffet}.{\BBCQ}
\newblock
\APACjournalVolNumPages{Journal of Fluid Mechanics}{850}{}{156--178}.
\PrintBackRefs{\CurrentBib}

\bibitem [\protect \citeauthoryear {%
Han%
\ \protect \BOthers {.}}{%
Han%
\ \protect \BOthers {.}}{%
{\protect \APACyear {2022}}%
}]{%
han2022numerical}
\APACinsertmetastar {%
han2022numerical}%
\begin{APACrefauthors}%
Han, B.%
, Xu, M.%
, Chen, G.%
, Cai, T.%
, Wang, W.%
\BCBL {}\ \BBA {} Guan, S.%
\end{APACrefauthors}%
\unskip\
\newblock
\APACrefYearMonthDay{2022}{}{}.
\newblock
{\BBOQ}\APACrefatitle {Numerical investigation of transonic buffet on a
  prescribed-pitching OAT15A airfoil} {Numerical investigation of transonic
  buffet on a prescribed-pitching oat15a airfoil}.{\BBCQ}
\newblock
\APACjournalVolNumPages{AIP Advances}{12}{3}{}.
\PrintBackRefs{\CurrentBib}

\bibitem [\protect \citeauthoryear {%
Hoffmann%
\ \BBA {} Chiang%
}{%
Hoffmann%
\ \BBA {} Chiang%
}{%
{\protect \APACyear {2000}}%
}]{%
hoffmann2000computational}
\APACinsertmetastar {%
hoffmann2000computational}%
\begin{APACrefauthors}%
Hoffmann, K\BPBI A.%
\BCBT {}\ \BBA {} Chiang, S\BPBI T.%
\end{APACrefauthors}%
\unskip\
\newblock
\APACrefYearMonthDay{2000}{}{}.
\newblock
{\BBOQ}\APACrefatitle {Computational fluid dynamics volume I} {Computational
  fluid dynamics volume i}.{\BBCQ}
\newblock
\APACjournalVolNumPages{Engineering education system}{}{}{}.
\PrintBackRefs{\CurrentBib}

\bibitem [\protect \citeauthoryear {%
Jameson%
, Schmidt%
\BCBL {}\ \BBA {} Turkel%
}{%
Jameson%
\ \protect \BOthers {.}}{%
{\protect \APACyear {1981}}%
}]{%
jameson1981numerical}
\APACinsertmetastar {%
jameson1981numerical}%
\begin{APACrefauthors}%
Jameson, A.%
, Schmidt, W.%
\BCBL {}\ \BBA {} Turkel, E.%
\end{APACrefauthors}%
\unskip\
\newblock
\APACrefYearMonthDay{1981}{}{}.
\newblock
{\BBOQ}\APACrefatitle {Numerical solution of the Euler equations by finite
  volume methods using Runge Kutta time stepping schemes} {Numerical solution
  of the euler equations by finite volume methods using runge kutta time
  stepping schemes}.{\BBCQ}
\newblock
\BIn{} \APACrefbtitle {14th fluid and plasma dynamics conference} {14th fluid
  and plasma dynamics conference}\ (\BPG~1259).
\PrintBackRefs{\CurrentBib}

\bibitem [\protect \citeauthoryear {%
Nair%
\ \BBA {} Sengupta%
}{%
Nair%
\ \BBA {} Sengupta%
}{%
{\protect \APACyear {1998}}%
}]{%
nair1998orthogonal}
\APACinsertmetastar {%
nair1998orthogonal}%
\begin{APACrefauthors}%
Nair, M\BPBI T.%
\BCBT {}\ \BBA {} Sengupta, T.%
\end{APACrefauthors}%
\unskip\
\newblock
\APACrefYearMonthDay{1998}{}{}.
\newblock
{\BBOQ}\APACrefatitle {Orthogonal grid generation for Navier--Stokes
  computations} {Orthogonal grid generation for navier--stokes
  computations}.{\BBCQ}
\newblock
\APACjournalVolNumPages{International journal for numerical methods in
  fluids}{28}{2}{215--224}.
\PrintBackRefs{\CurrentBib}

\bibitem [\protect \citeauthoryear {%
Sengupta%
}{%
Sengupta%
}{%
{\protect \APACyear {2013}}%
}]{%
sengupta2013high}
\APACinsertmetastar {%
sengupta2013high}%
\begin{APACrefauthors}%
Sengupta, T.%
\end{APACrefauthors}%
\unskip\
\newblock
\APACrefYear{2013}.
\newblock
\APACrefbtitle {High accuracy computing methods: fluid flows and wave
  phenomena} {High accuracy computing methods: fluid flows and wave phenomena}.
\newblock
\APACaddressPublisher{}{Cambridge University Press}.
\PrintBackRefs{\CurrentBib}

\end{thebibliography}

\end{document}